# Multi-dimensional vibration sensing and simultaneous self-homodyne optical transmission of single wavelength net 5.36 Tb/s signal using telecom 7-core fiber

**Jianwei Tang[1,2], Xueyang Li[*,1], Bang Yang[2], Chen Cheng[2], Yaguang Hao[1,2], Yifan Xu[1], Jiali Li[1], Zhixue He[1], Yanfu Yang[*,1,2], and Weisheng Hu[1]**

[1]*Department of Circuits and System, Peng Cheng Laboratory, Shenzhen, China*
[2]*School of Electronics and Information Engineering, Harbin Institute of Technology, Shenzhen, China*
E-mail: xueyang.li@pcl.ac.cn; yangyanfu@hit.edu.cn

**Abstract:** We present a high-capacity self-homodyne optical transmission system that enables simultaneously multidimensional vibration sensing based on a weakly-coupled 7-core fiber. To our knowledge, we demonstrate for the first-time detection of fiber vibration direction along with strength, frequency, and location of the vibration source, while transmitting in the meantime single-carrier 16 QAM signal reaching a net date rate of 5.36 Tb/s over 41.4 km of telecom 7-core fiber.

## 1. Introduction

The ever-increasing demand for higher transmission capacity attracts growing interests on space-division multiplexing (SDM) fiber technologies. With the capability for ultra-high spatial multiplicity over 100 per fiber [1], SDM fiber shows the potential in overcoming the per-fiber capacity of conventional single-mode fiber (SMF). Amongst different types of SDM fibers, weakly-coupled (WC) multi-core fiber (MCF) will be the first to enter commercial use in submarine network [2]. Transforming the MCF infrastructure into a distributed sensor network offers important opportunities for applications in smart city and enable early warning of disasters such as earthquakes and tsunami.

Simultaneous transmission of SDM signals and distributed acoustic sensing (DAS) has been successfully realized within a single WC-MCF fiber harnessing the unique feature of independent spatial channels with minimal crosstalk [3-4]. Compared to DAS, the forward vibration sensing scheme based on phase detection offers distinct advantages including an extended sensing range and compatibility with existing fiber telecom infrastructure equipped with isolators in amplifier modules. Prior research has validated the effectiveness of the phase-based forward vibration sensing in locating the vibration event *in conventional SMF* by correlating the phases demodulated from either two counter-propagating continuous wave (CW) carriers or communication signals [5-6]. Due to the proportionality between the fiber longitudinal strain and the phase variation, precise detection of vibration frequency and strength are achievable. However, simultaneous optical communications and forward vibration sensing based on phase detection using WC-MCF remain unexplored, and it is desirable to enhance the capabilities of vibration detection by exploring the core layout of WC-MCF.

In this paper, we present a novel forward vibration detection scheme that can be readily integrated with a SDM self-homodyne transmission system utilizing a weakly-coupled 7-core fiber. Our approach retrieves the phases of a remotely sent local oscillator (LO) in a dedicated fiber core and a counter-propagating continuously-wave (CW) carrier in a core used for telecom signal, yet without affecting the communication performance. Our scheme realizes multi-dimensional vibration sensing, enabling detection of strength, frequency, location, and most importantly, direction of vibration. In addition, our scheme harnesses the advantages of self-homodyne detection [7], eliminating the need of carrier phase recovery and enabling transmission a net data rate of 5.36 Tb/s over 41.4 km of 7-core fiber.

## 2. Working Principle

Fig. 1 illustrates the principle of the proposed scheme. In the presence of directional fiber vibration, the changes of the fiber core length $\Delta L$ are not equal at each instant of vibration due to heterogeneously distributed longitudinal strain in each fiber core. $\Delta L$ depends on the location of each fiber core as well as the direction of vibration. We define in Fig. 1(a) angle $\theta$ between the vibration direction and the reference axis that is normal to the plane formed by core1 and core7. With continuous-wave carrier 1 (CW1) injected to core7 and a counter-propagating continuous-wave carrier 2 (CW2) injected to core1, respectively, the phase variations $\Phi_1(t)$ and $\Phi_2(t)$ encoded in CW1 and CW2 due to fiber vibration diverge from each other except in critical cases where $\theta$ is 0º or 180º. This phase divergence is due to different optical length variations induced in core1 and core7. Thus, we can detect the direction of vibration by processing the differential-mode phase information of $\Phi_1(t)$ and $\Phi_2(t)$. In our scheme, 6 branches of telecommunication signals are forward propagating in *core1 to core6*, respectively. Self-homodyne detection is realized by exploiting CW1 in core7. Note that by optimizing the power of CW2, the influence of its Rayleigh backscatter on the telecom signal in core1 is marginal.

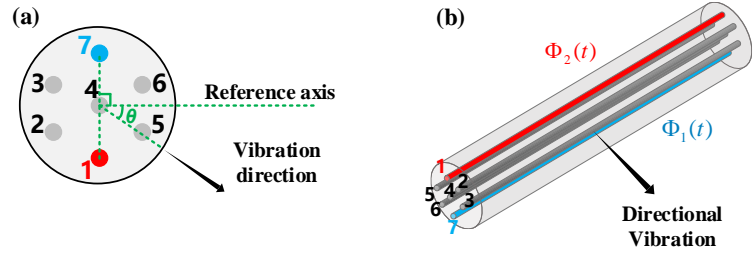

Fig. 1 (a) Cross-section of 7-core fiber. (b) Schematic of the proposed MCF-based acoustic sensing scheme with integrated self-homodyne communications.

For a given fiber vibration at position Z, the phase disturbance in CW1 travels over a length of *L-Z*, whereas the phase disturbance of CW2 propagates over a fiber length of *Z*. The two CW carriers are received by two coherent receivers which achieve timing synchrony, for example, by GPS. The vibration location Z can be detected based on the equation below:

$$Z = \frac{L}{2} - \frac{\Delta t \cdot c}{2n} \quad (1)$$

where $\Delta t$ is the time delay between the phases from two coherent receivers, $c$ is the speed of light, and $n$ is the group index.

## 3. Experimental setup

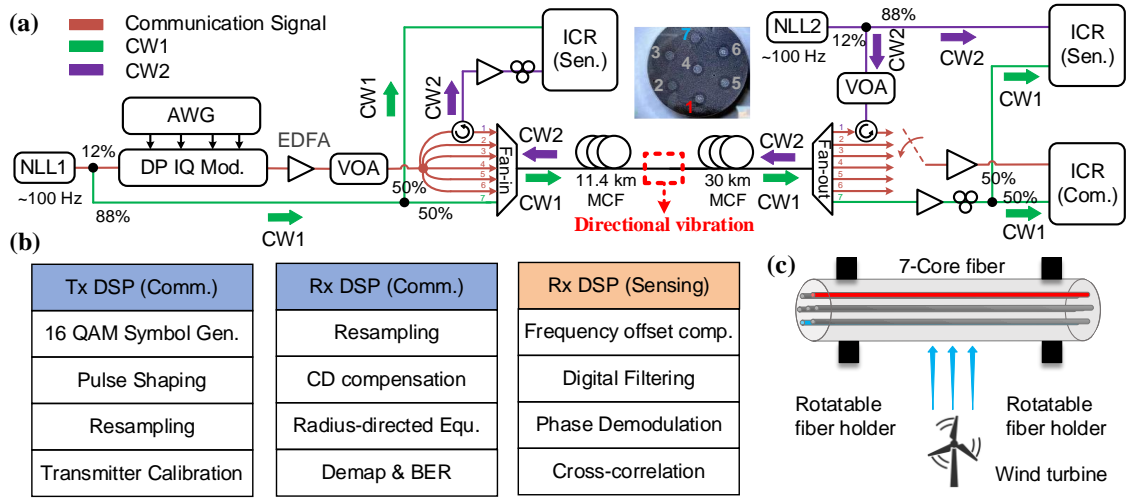

Fig. 2 (a) Experimental setup. (b) Tx DSP, Rx DSP for communication, and Rx DSP for sensing. (c) Generation of directional vibration.

In this section, we experimentally demonstrate the simultaneous communication and acoustic sensing scheme using a weakly-coupled 7-core multi-core fiber (MCF) with inter-core crosstalk below -40 dB. Two narrow linewidth lasers (NLLs, NKT X15) at 1550 nm with an output power of 16 dBm are used as optical sources for both communication and sensing. The frequency offset between the two NLLs is ~5 MHz. The electrical signals generated by a 128 GSa/s arbitrary waveform generator (AWG, Keysight 8199A) drive a 4-channel dual-polarization IQ modulator (DP-IQM). 12% of optical power from NLL1 is fed to the DP-IQM as the optical carrier. The modulated signal is amplified, split into six branches, and coupled to 6 cores of the MCF, respectively, via a fan-in module in order to emulate the transmission of 6 independent optical signals. The remaining 88% of optical power of NLL1 is equally split into two parts. One part serves as the co-traveling LO (CW1) that is coupled into core7 and sent to the receiver for self-homodyne coherent detection (SHCD) of telecom signals. The other part is used to beat with the counter-propagating sensing CW carrier in core1 at Tx.

At the receiver end, CW1 transmitted in core 7 is amplified by an EDFA and subsequently aligned to a desired polarization state for equal power splitting. Next, CW1 is divided into two branches, which is used as the LO for SHCD of communication signals and for distributed vibration sensing, respectively. The optical signals in core 1~6 are amplified and received using an ICR, and then sampled by a real-time oscilloscope (RTO, Keysight UXR0594AP) operating at 256 GSa/s. Next, 12% optical power of NLL2 is coupled back into core1 via a circulator. The variable optical attenuator (VOA) is used to optimize the power of CW2 so that the impact of Rayleigh scattering on the telecom signal is negligible. The rest 88% of optical power is used as the LO for heterodyne detection of CW1 transmitted in core7. The DSP process of the transmitter and receiver is also shown in the Fig. 2(b). Based on the SHCD communication system using 7-core MCF, vibration sensing can be realized simultaneously through a simple DSP scheme.

Directional vibration is induced in a short section of MCF by use of a wind turbine as shown in Fig. 2(c). The fiber section is clamped by a pair of rotatable fiber holders in order to precisely control the angle $\theta$ as defined earlier.

## 3. Results and Discussions

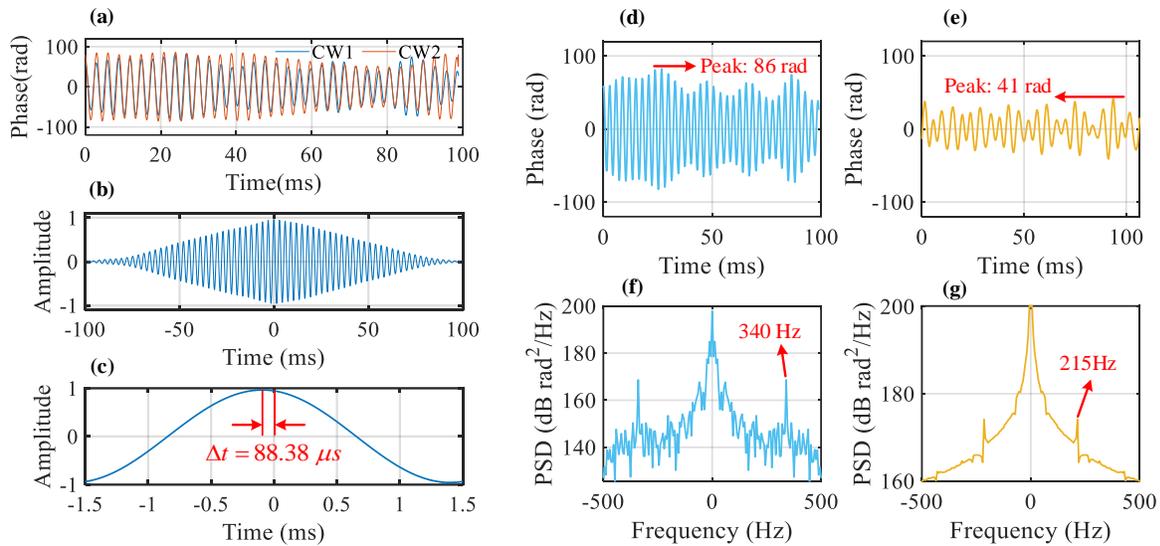

Fig. 3 (a-c) are shown for vibration location: (a) Phase waveforms of CW1 and CW2 versus time. (b) Cross-correlation between the phases of CW1 and CW2. (c) The enlarged view of the central area of (b). (d-g) are shown for vibration strength and frequency sensing: (d) and (e) are the recovered phases of CW1 when the wind turbine is working at mode H and mode L respectively. (f) and (g) are the corresponding PSD of phases obtained at mode H and mode L, respectively.

*A. Vibration Location*

Fig. 3(a) depicts the demodulated phase waveforms of the forward and backward propagating CW carriers, i.e. CW1 and CW2, respectively. It is found that the phase changes of the CW carriers have a near-sinusoidal shape. In order to locate the vibration, we cross-correlate the phase waveforms of CW1 and CW2 as depicted in Fig. 3(b). Fig. 3(c) shows the enlarged view of the central area of the cross-correlation, which reveals that the peak position corresponds to a time interval of 88.38 $\mu s$, corresponding to a position of 11.65 km according to Eq. (1), which is in agreement with the expected vibration location of 11.4 km (measured by $\varphi$-OTDR) taking into account the the EDFA, PC, circulator and fiber patch cords in the system.

*B. Vibration Strength and Frequency Sensing*

In order to investigate the phase impact of the vibration strength, we detect the phase of CW1 by setting the wind turbine either in a high-power mode (mode H) or a low-power mode (mode L). Fig. 3(d-e) shows the phase waveforms of CW1 at the two aforementioned conditions, respectively.

It can be seen from the figure that the maximum phase changes are ~85 radians and ~40 radians at mode H and mode L, respectively, which is in agreement with the perceived amplitude of vibration. The change of vibration frequency at the two modes of the wind turbine is also observed. Fig. 3(f-g) show the power spectral density (PSD) of the phases in mode H and mode L, respectively. As can be seen from Fig. 4(f-g), mode H induces a higher vibration frequency of 340 Hz, whereas mode L induces a lower vibration frequency of 215 Hz. The results validated the effectiveness of our scheme in detecting the strength and frequency of vibration.

*C. Detection of Vibration Direction*

Fig. 4(a) and (b) plot the phases as a function of time when angle $\theta$ is 0° and 90°, respectively, for CW1 and CW2. We also show the phase difference between CW1 and CW2 in Fig. 4(c) at the two corresponding angles of $\theta$. We observe more significant waveform difference between CW1 and CW2 at $\theta$=90° compared to the critical case where $\theta$=0°. The stronger phase difference when $\theta$ is 90° is due to more divergent longitudinal strain induced in core1 and core7, respectively. Next, we tune angle $\theta$ and calculate the average power of the differential phase (APDP) between CW1 and CW2 as shown in Fig. 4(d). The change of the APDP versus $\theta$ is in agreement with our earlier analysis. Note that the ambiguity in determining the angle $\theta$ can be resolved by counter-propagating CW carriers in more telecom spatial channels.

*D. Optical transmission of 120 Gbaud DP-16QAM signal*

We show the BER of the 120 Gbaud DP-16QAM signal propagating in core 1~6 of the 7-core fiber in the Fig. 4(e). As shown in this figure, the impact of the Rayleigh backscatter from CW2 on the telecom data in core1 is negligible. The constellation diagrams obtained in core1 and the central core 4 are shown in the inset. We transmit 6×960 Gbps 16QAM signal over 41.4 km of WC 7 core fiber and reaches a BER below the threshold of 3.8E-3. By excluding the FEC overhead of 7%, 5.36 Tbps net data rate is achieved.

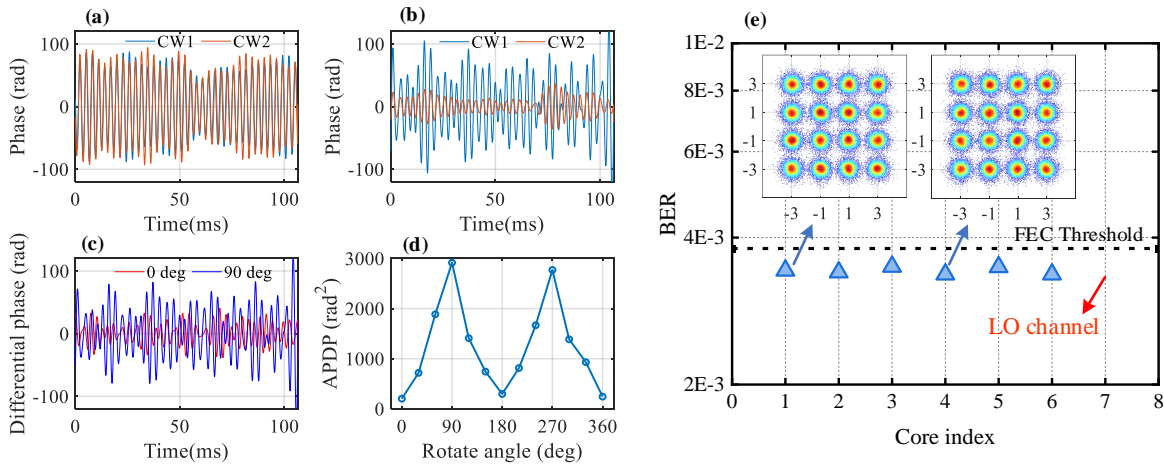

Fig. 4 Phase waveforms of CW1 and CW2 when (a) $\theta$=0° and (b) $\theta$=90°, respectively. (c) The phase difference between CW1 and CW2 when $\theta$=0° and $\theta$=90°. (d) Average power of the phase difference (ADPD) versus $\theta$.

## 4. Conclusion

In this paper, we propose an integrated self-homodyne optical transmission system that simultaneously achieves multidimensional vibration sensing based on a WC 7-core fiber. Leveraging the differential phase information retrieved from two-counter-propagating CW carriers in a MCF, we show for the first time distributed detection of strength, frequency, direction, as well as location of vibration, while transmitting in the meantime dual-pol SDM 16 QAM signals having a net data rate of 5.36 Tbps. The high-capacity SDM self-homodyne communication system integrated with multi-dimensional vibration sensing capability offers a cost-effective and multi-functional candidate for the marine activities monitoring, early warning of tsunami and earthquakes.

**Acknowledgments**

This work is supported by the National Key R&D Program of China (2020YFB1806401), the National Natural Science Foundation of China (62201308), and the Science Technology and Innovation Commission of Shenzhen Municipality (JCYJ20210324131408023).

**Data availability**

Data underlying the results presented in this paper are not publicly available at this time but may be obtained from the authors upon reasonable request.

**Conflict of interest**

The authors declare that they have no conflict of interest.